\documentclass[conference]{IEEEtran}
\IEEEoverridecommandlockouts
\usepackage{cite}
\usepackage{amsmath,amssymb,amsfonts}
\usepackage{algorithmic}
\usepackage{graphicx}
\usepackage{textcomp}
\usepackage{xcolor}
\usepackage{siunitx}
\usepackage{booktabs} 
\usepackage{multirow}
\usepackage{booktabs,threeparttable}
\ifCLASSOPTIONcompsoc
    \usepackage[caption=false, font=normalsize, labelfont=sf, textfont=sf]{subfig}
\else
\usepackage[caption=false, font=footnotesize]{subfig}
\fi

\usepackage{siunitx}
\def\BibTeX{{\rm B\kern-.05em{\sc i\kern-.025em b}\kern-.08em
    T\kern-.1667em\lower.7ex\hbox{E}\kern-.125emX}}
\begin{document}

\title{Characterization of Spatial-Temporal Channel Statistics from  Indoor Measurement Data at D Band\\
\author{\IEEEauthorblockN{Chathuri~Weragama$^\star$,~Joonas~Kokkoniemi$^\star$,~Mar~Francis~De~Guzman$^\dagger$,\\~Katsuyuki~Haneda$^\dagger$,~Pekka~Ky\"osti$^\star$~and~Markku~Juntti$^\star$}
\IEEEauthorblockA{$^\star$Centre for Wireless Communications (CWC), University of Oulu, P.O. Box 4500, 90014 Oulu, Finland}
\IEEEauthorblockA{$^\dagger$Department of Electronics and Nanoengineering, Aalto University, 02150 Espoo, Finland}
}
}

\maketitle

\begin{abstract}
Millimeter-wave (mmWave) and D Band (110--170~GHz) frequencies are poised to play a pivotal role in the advancement of sixth-generation (6G) systems and beyond, owing to their ability to enhance performance metrics such as capacity, ultra-low latency, and spectral efficiency. This paper concentrates on deriving statistical insights into power, delay, and the number of paths based on measurements conducted across four distinct locations at a center frequency of 143.1 GHz. The findings underscore the suitability of various distributions in characterizing power behavior in line-of-sight (LOS) scenarios, including lognormal, Nakagami, gamma, and beta distributions, whereas the loglogistic distribution gives the optimal fit for power distribution in non-line-of-sight (NLOS) scenarios. Moreover, the exponential distribution shows to be the most appropriate model for the delay distribution in both LOS and NLOS scenarios. In terms of the number of paths, observations indicate a tendency for the highest concentration within the 10 m to 30 m distance range between the transmitter (Tx) and receiver (Rx). These insights shed light on the statistical nature of D band propagation characteristics, which are vital for informing the design and optimization of future 6G communication systems
\end{abstract}


\section{Introduction}

In the past decade, exploration into millimeter wave (mmWave) frequencies up to 100 GHz has paved the way for their integration into fifth-generation communication systems, offering enhanced performance metrics including increased capacity, ultra-low latency, and spectral efficiency, among others \cite{dou201545ghz,solomitckii2018characterization, naderpour2016spatio}. However, the evolving demands of sixth-generation (6G) communication, marked by higher data rates, seamless connectivity, and greater device density to support ultra-reliable low-latency communication (URLLC), massive machine-type communications (mMTC), and enhanced mobile broadband (eMBB), necessitate a deeper investigation into propagation and channel characteristics at even higher frequencies. The D band, spanning from \qty{110} {GHz} to \qty{170}{GHz}, has emerged as a focal point for addressing these requirements in 6G. Notably, recent studies \cite{9611219, wang2022reflection} have delved into material characterization and channel modeling at the D band, particularly in indoor and outdoor environments, considering reflection, diffraction, and penetration losses. While existing research has endeavored to characterize path loss, angular spread, and inter-cluster characteristics at the D band using indoor measurements \cite{7095539,8936412},  challenges persist due to high scattering, attenuation, and interference caused by the short wavelengths, rendering the prediction of channel characteristics a formidable task.

In this work, we aim to develop a statistical model for indoor D band communication, drawing insights from measurements conducted at three distinct locations: Sello shopping mall, Helsinki-Vantaa Airport (operating at a center frequency of 143.1 GHz), and the entrance hall of Aalto University (operating at a center frequency of 142 GHz). Notably, measurements at the Aalto University entrance hall were collected during two separate time periods, treated as distinct locations for analytical purposes. Our primary objective is to formulate a statistical model with broad applicability to indoor communications in the D band, while discerning the channel behavior in both line-of-sight (LOS) and non-line-of-sight (NLOS) scenarios. The collected data undergoes comparisons against theoretical distributions to elucidate their inherent properties. Specifically, our analysis focuses on delay, power, and the number of paths, while our future work entails deriving stochastic multiple-input multiple-output (MIMO) channel models by leveraging the statistics obtained from the measurements. Our aim is to aggregate all measured paths within specific locations to derive statistical distributions for amplitude or path loss, delay, the number of distinguishable paths, and the angle of arrival and departure. The  distributions, combined with a predetermined antenna array configuration, will make it possible to generate random MIMO channel power-delay-angle profiles by arbitrarily summing paths from the derived distributions. 

This paper is organized as follows. Section \ref{sec:mesaure} provides an overview of the data measurement procedure and related work, while Section \ref{sec:methodology} delineates the methodology employed for data processing and analysis of statistical parameters. Section \ref{sec:results} presents the findings of our analysis and associated statistical metrics. Finally, Section \ref{sec:conclusion} encapsulates the conclusions drawn from our study.

\section{Measurement Data and Related Works} \label{sec:mesaure}

The channel measurements on D band were conducted by Aalto university in various locations in Helsinki region, Finland. Aalto University have released several datasets from these measurements, which are openly available in \cite{de_guzman_2023_7640353}. While there are multiple datasets in outdoor locations, in this work we focus on the indoor ones. The chosen indoor locations comprise commercial buildings: Sello shopping mall, Helsinki-Vantaa Airport, and the entrance hall of Aalto University. All of these location have relatively open floor plan, but they also have notable differences. The shopping mall is a multi-storey building with open areas but also more closed areas. The airport is mostly a single floor, but very open, and the Aalto university is partially open space, but with more obstacles and features. The exact measurement setups have been detailed in various previous works, such as in \cite{DeGuzman2022entrance} for the Aalto University entrance hall, \cite{nguyen2018shopping} for the shopping mall, and \cite{nguyen2021airport} for the airport and shopping mall. 
The shopping mall and airport measurements were conducted at \qty{143.1}{GHz} center frequency, while Aalto University entrance hall measurement frequency was \qty{142}{GHz} with \qty{4}{GHz} bandwidth in all locations. The transmitter (Tx) end was biconical omnidirectional antenna (0 dBi) and the receiver (Rx) end was a horn antenna on rotating platform (19 dBi). The radio frequency (RF) power was about \qty{-7}{dBm}.


There are several ways to form channel models out of experimental data. Popular methods include, for instance, close-in (CI), floating intercept (FI), and dual slope methods \cite{Sooyoung2016mmWave, Youngbin2014dual, Tataria2020channel, Haneda2016}. These methods are based on fitting a path loss curve over the measurement data and using a random variable to model fading/shadowing. The random variable tends to follow either normal (Gaussian) or lognormal distribution. These models are very handy when calculating the expected path loss in a certain environment. The methods vary in accuracy depending on the distance and the environment. In principle, it is possible to use them to get very reliable path loss values. We aim at deriving the statistics of the channels with slightly different approach, where we analyse the characteristics of the multipath components of the channel in order to estimate the channel statistics. Our ultimate goal is to derive stochastic MIMO channel models by utilizing the statistics obtained from the measurements. While a similar approach to MIMO channel modeling has been undertaken in \cite{Das2024mimo}, it is noteworthy that our statistics are derived solely from pure measurement data.

The same measurement data has been analysed before with respect to the fading statistics \cite{Papasotiriou2021fading }. Our work here differs from that by aggregating the data per location in order to obtain statistically independent distributions as any single measurement link is strongly related to the environment geometry. When aggregating all the measurement positions per measurement location, we will obtain a stochastic data set as possible given the limitation of the real world measurements. The process of obtaining the statistics is described in the next section and the actual channel modeling process will be given in the future work.

\section{Modelling Methodology} \label{sec:methodology}

The data collection process, discussed in Section \ref{sec:mesaure}, involved aggregating data concerning multipath components from a single locations into a unified dataset. Subsequently, we conducted an analysis of the power delay profile (PDP), power distribution, delay distribution, and the number of paths for both LOS and NLOS scenarios. 
Following this organization, the data underwent normalization to mitigate the influence of distance and to examine power and delay distributions. Normalized power, adjusted to compensate for free space path loss (FSPL) and obtain only the impact of the channel on the received power is expressed as
\begin{equation}\label{equ:power_norm}
P_{n(dB)} = P_{(dB)} - 20\log_{10}\left(\frac{4\pi f d}{c}\right),
\end{equation}
where $P_{n(dB)}$ denotes the normalized power, $P_{(dB)}$ is the measured power, and $d$, $f$, and $c$ represent the length of the path, operating frequency, and speed of light in free space respectively. The delay, on the other hand, is normalized with the delay of the first arrival in order to model the delay spread with the distributions. The normalization of delay was calculated as
\begin{equation}
\label{equ: delay_norm}
\tau_{n} = \tau - \tau_{L},
\end{equation}
where $\tau_{n}$, $\tau$, and $\tau_{L}$ correspond to the normalized delay, measured delay, and the delay of the first-arrived path, respectively. Modeling the number of paths posed challenges due to inherent sparsity in the measurements. However, despite these challenges, we were able to derive reasonably accurate statistics for the number of paths as well.

We conducted an analysis of normalized power and delay data to elucidate the behavior of power and delay profiles within the D band. We employed various metrics to assess the adequacy of fit between empirical and theoretical distributions of the measured data. Specifically, we utilized the Kolmogorov–Smirnov test (KS test) \cite{massey1951kolmogorov}, the Kullback–Leibler (KL) divergence \cite{kullback1951information}, the quantile-to-quantile (Q-Q plots), and the Anderson–Darling test. Due to the limited number of samples and the diverse range of theoretical distributions under consideration, we focused our evaluation primarily on the KS test and Q-Q plots. The KS test determines the goodness of fit by comparing the maximum distance between the empirical cumulative distribution function (CDF) and the theoretical CDF. A p-value exceeding the significance threshold indicates a satisfactory fit between the two distributions. Conversely, Q-Q plots compare the quantiles of the first dataset against those of the second dataset. A perfect match between datasets from the same distribution would result in a plot following a $y=x$ line. The correlation coefficient of this plot serves as a metric for assessing the congruence between the two distributions. 

The analysis has been done in a Python environment using libraries numpy, pandas, scipy.stats for data manipulation, data organizing and statistical fitting respectively. The main tools which aided the visual representations of the said analysis is matplotlib and seaborn libraries.

\section{Results}\label{sec:results}
We have analyzed the measurement multipath data based on location to understand the behaviour of power, delay and multipath propagation at D band. The number of data points in each location is listed in Table \ref{tab:Number of Data Points} and as per discussed in the previous section. The Sello, Airport, TUAS, and TUAS2 locations refer to Sello shopping mall,  Helsinki-Vantaa Airport, initial measurement campaign at the entrance hall of Aalto University and the second measurements respectively. The results of these analysis is discussed in this section.

 \begin{table}[t]
\caption{Number of Data Points}
    \label{tab:Number of Data Points}
    \centering
    \begin{tabular}{l |c| c} \hline
    Location &     LOS &  NLOS\\ \hline
       Sello &   304 & 29\\ 
       Airport &   375 & 41\\ 
       TUAS &   29 & 387\\ 
       TUAS2 &   268 & 1812\\ \hline
       Total &   986 & 2269\\ 
       
       \hline
    \end{tabular}
    
\end{table}

\subsection{Power Delay Profile }

\begin{figure} [t]
    \centering
  \subfloat[Sello\label{1a}]{%
       \includegraphics[width=0.5\linewidth]{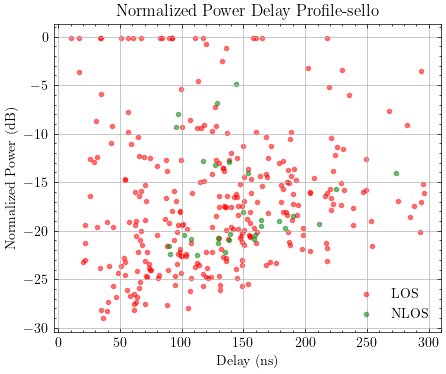}}
    \hfill
  \subfloat[Airport\label{1b}]{%
        \includegraphics[width=0.5\linewidth]{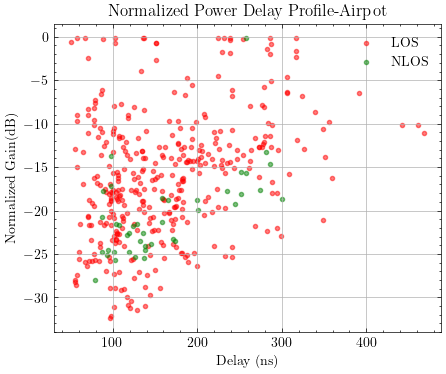}}
    \\
  \subfloat[TUAS\label{1c}]{%
        \includegraphics[width=0.5\linewidth]{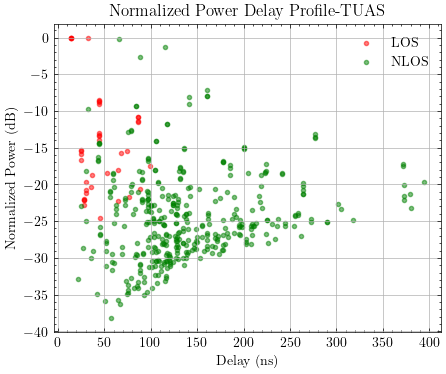}}
    \hfill
  \subfloat[TUAS2\label{1d}]{%
        \includegraphics[width=0.5\linewidth]{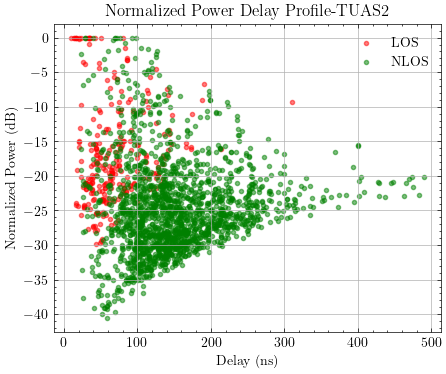}}
  \caption{Power Delay Profiles}
  \label{fig:PDP} 
\end{figure}

To visualize and analyze the power delay profile (PDP) of the measured data, we have normalized the measured power data as presented in (\ref{equ:power_norm}) and visualized it against the measured delay to understand the PDP, eliminating the effect of FSPL. Figure \ref{fig:PDP} represents the PDP of all four location measurements for both LOS and NLOS scenarios.

Upon analyzing the PDP visualizations, it becomes apparent that a significant number of data points exhibit zero power after normalization. This occurrence is likely attributed to FSPL, indicating that these data points likely correspond to LOS paths between Tx and Rx. Interestingly, it is noteworthy that not only LOS data points but also certain NLOS data points exhibit zero normalized power, implying the presence of LOS paths even in NLOS scenarios. These could be caused by reflected signals from highly reflective objects. Furthermore, an observed lower bound in the PDP of TUAS and TUAS2 locations may be attributed to a reduced dynamic range in multipath components due to noise interference. This lower bound is specifically discernible in the TUAS and TUAS2 locations, likely owing to the availability of a sufficient number of data points for observation, whereas other locations may lack an adequate dataset for such analysis.

\subsection{Normalized Power Distribution (NPD)}

\begin{figure} [t]
    \centering
  \subfloat[Sello]{%
       \includegraphics[width=0.5\linewidth]{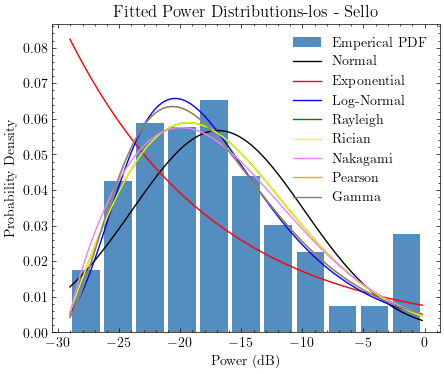}}
    \hfill
  \subfloat[Airport]{%
        \includegraphics[width=0.5\linewidth]{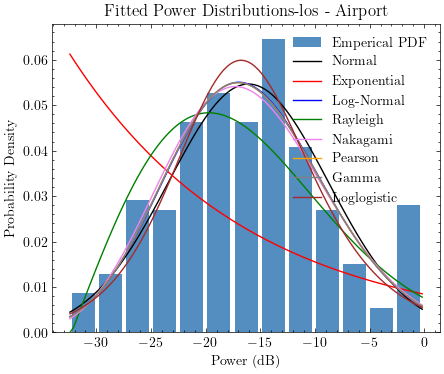}}
    \\
  \subfloat[TUAS]{%
        \includegraphics[width=0.5\linewidth]{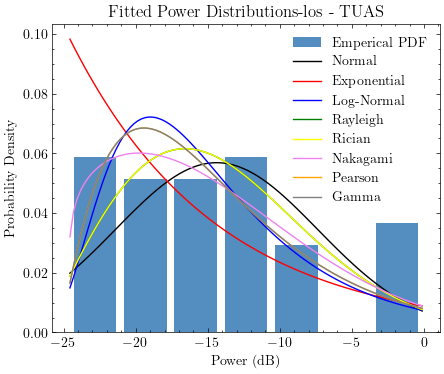}}
    \hfill
  \subfloat[TUAS2]{%
        \includegraphics[width=0.5\linewidth]{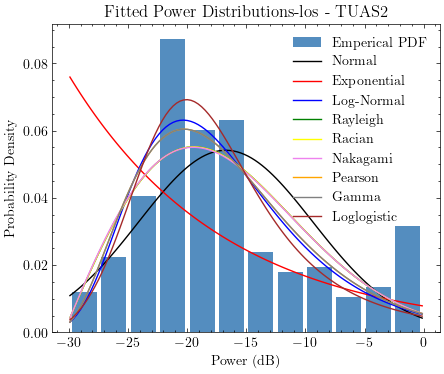}}
  \caption{Normalized Power Distribution - Line-of-Sight}
  \label{fig:NPD - LOS} 
\end{figure}

\begin{figure} [t]
    \centering
  \subfloat[Sello]{%
       \includegraphics[width=0.5\linewidth]{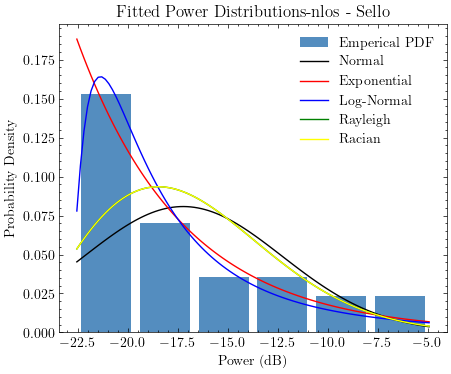}}
    \hfill
  \subfloat[Airport]{%
        \includegraphics[width=0.5\linewidth]{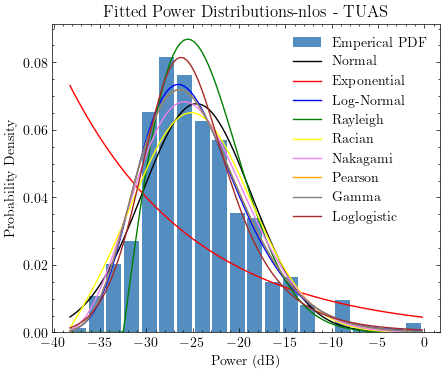}}
    \\
  \subfloat[TUAS]{%
        \includegraphics[width=0.5\linewidth]{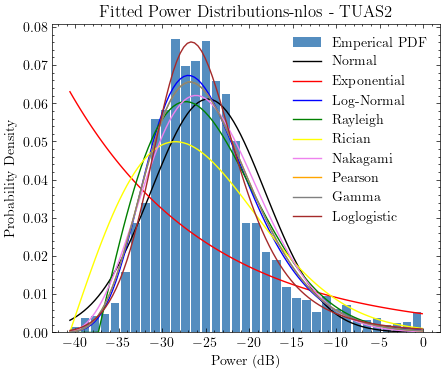}}
    \hfill
  \subfloat[TUAS2]{%
        \includegraphics[width=0.5\linewidth]{Results/Power_Distribution/NLOS/Fitted_Power_Distributions-Nlos-TUAS2.png}}
  \caption{Power Delay Profiles - Non-Line-of-Sight}
  \label{fig:NPD - NLOS} 
\end{figure}

To perceive the distribution of power of multipath components at D band we have built the empirical probability density function (PDF) of the multipath components of a single location and attempted fitting different distributions,  considering both LOS and NLOS scenarios. Figure \ref{fig:NPD - LOS} and \ref{fig:NPD - NLOS} provides the empirical and theoretical distributions related to LOS and NLOS respectively. The parameters related to fitting, as well as the measurements of the KS-test and Q-Q plots regarding the goodness of fit, are presented in Table \ref{tab:NPD - Results - Summary}. The loc, scale, and shape columns specified in Table \ref{tab:NPD - Results - Summary} represent parameters related to each distribution fitting, estimating the location, scale, and shape of the distribution, respectively.

\begin{table*} [t]
    \centering
    \caption{NPD - Results - Summary}
    \label{tab:NPD - Results - Summary}
    \begin{tabular}{l |l| ccccc| ccccc}
    \hline
    \multirow{2}{*}{Location} &
    \multirow{2}{*}{Distribution} &
        \multicolumn{5}{c}{LOS} & 
        \multicolumn{5}{c}{NLOS} \\ 
    & &   p-value & R & Loc &  Scale & Shape &  p-value & R &  Loc &  Scale & Shape\\ \hline
    
   \multirow{9}{*}{Sello} & Normal      & 0.0325 & 0.97 & -17  & 7 & -  & 0.0957 & 0.92 & -17.2 & 4.93  & - \\
    & Exponential & \num{9.4e-12} & 0.97 &  -29 &  12.1 & -  & 0.8664 & 0.96 & -22.5  & 5.31  & -  \\
    & Log-Normal  & 0.8526 & 0.99 & -35.5  & 17.4  & 0.37  & 0.7512 & 0.91 & -23.1  & 4.03  & 0.91  \\
    & Rayleigh  & 0.4456 & 0.99 & -29.6  & 10.3  & - & 0.0741 & 0.96 & -24.95  & 6.48  & -  \\
    & Rician & 0.4456 & 0.99 & -29.6  & 10.3 & 0  & 0.0741 & 0.96 & -24.95  & 6.48 & 0 \\
    & Nakagami  & 0.5238 & 0.99 & -29.3   & 14.2  & 0.876  & 0.7722  & 0.98 & -22.53  & 7.26 & 0.88  \\
    & Gamma  & 0.8106 & 0.99 &-30.7  & 3.73  & 3.66  & 0.7210 & 0.98 & -22.53  & 5.74  & 0.88  \\
    & Beta & 0.8106 & 0.99 & -16.9& 7.14 & 1.04  & 0.8828 & 0.98 & -17.12   & 5.69  & 2.1  \\
    & Log Logistic  & \num{2.1e-34} & 0.57 &  -29   & 5.60  & 0.84  & \num{5.546e-05} & 0.58 & -22.53 & 1.04  & 0.83  \\ \hline
    
    \multirow{9}{*}{Airport} & Normal      & 0.6897 & 0.99 & -16.08  & 7.3 & -  & 0.5092& 0.92 & -20.56 & 4.83  & - \\
    & Exponential & \num{2.51e-24} & 0.93 &  -32.42 &  16.33 & -  & 0.0141 & 0.97 & -27.95  & 7.38  & -  \\
    & Log-Normal  & 0.8837 & 0.99 & -105.55 & 89.16  & 0.08 & 0.8920 & 0.97 & -30.97  & 9.48 & 0.42  \\
    & Rayleigh  & 0.0057 & 0.99 & -32.27  & 16.3  & - & 0.7153 & 0.96 & -28.75  & 6.72  & -  \\
    & Rician & 0 & 0.27  & -32.4  & 2.21 & 0  & 0.7153 & 0.95 & -28.75  & 6.72 & 0  \\
    & Nakagami  & 0.8393 & 0.99 & -41.2   & 26.16  & 3.04 & 0.8640 & 0.96 & -28.23  & 9.06 & 0.79  \\
    & Gamma  & 0.8927 & 0.99 &-72.57  & 0.94 & 59.7 & \num{1.5e-31} & 0.97 & -27.95 & 1.55  & 0.73  \\
    & Beta & 0.8927 & 0.99 & -16 & 7.31 & 0.26  & 0.9718 & 0.97 & -20.56   & 4.55  & 1.08  \\
    & Log Logistic  & 0.72 & 0.99 &  -102.75   & 86.38  & 20.64  & \num{7.6e-05} & 0.78 & -27.95 & 4.14 & 0.88  \\ \hline
    
    \multirow{9}{*}{TUAS} & Normal      & 0.6357 & 0.95 & -14.38  & 7 & -  & 0.0269 & 0.97 & -24.57 & 5.88 & - \\
    & Exponential & 0.2978 & 0.93&  -24.56 &  10.17 & -  & \num{7.37e-36} & 0.90 & -38.24  & 13.67  & -  \\
    & Log-Normal  & 0.9650 & 0.94 & -28.21  & 12.12  & 0.52  & 0.9235 & 0.96 & -48.56  & 23.3  & 0.24  \\
    & Rayleigh  & 0.7257 & 0.97  & -26.43  & 9.85  & - & \num{3.06e-05} & 0.96 & -32.5  & 6.98  & -  \\
    & Rician & 0.7257 & 0.97 & -26.43  & 9.85 & 0  & 0.1117 & 0.96 & -38.54  & 6.6 & 1.8  \\
    & Nakagami  & 0.9070 & 0.97 & -24.67   & 12.44 & 2.16 & 0.3818 & 0.96 & -39.53  & 16.07  & 1.78  \\
    & Gamma  & 0.9581 & 0.96 &-25.27  & 5.04  & 1.36  & 0.8315 & 0.99 & -42.47  & 1.88  & 9.50  \\
    & Beta & 0.9581 & 0.96 & -14.39 & 7.41& 1.04  & 0.8315 & 0.99 & -24.57  & 5.8  & 0.64  \\
    & Log Logistic  & \num{2.37e-05} & 0.47 &  -24.56   & 3.38  & 0.81  & 0.9695
 & 0.99 & -46.11 & 20.78  & 6.62  \\ \hline
 
    \multirow{9}{*}{TUAS2} & Normal      & \num{3.47e-04} & 0.94 & -16.7  & 7.3 & -  & \num{9.88e-12}
 & 0.98 & -24.63 & 6.5  & - \\
    & Exponential & \num{7.06e-15} & 0.90 &  -29.9 &  13.17 & -  & \num{8.969 e-177} & 0.24 & -40.61  & 15.86  & -  \\
    & Log-Normal  & 0.2833 & 0.98 & -36.4  & 18.38  & 0.37  & \num{2.685e-03} & 0.98 & -50.15  & 24.62  & 0.24  \\
    & Rayleigh  & 0.0172 & 0.98 & -30.42  & 11  & - & \num{5.165e-06} & 0.98 & -37.34  & 10  & -  \\
    & Rician & 0.0172 & 0.98 & -30.42  & 11  & 0 & \num{2.920e-32} & 0.98 & -40.63  & 12.1 & 0.11  \\
    & Nakagami  & 0.0182 & 0.98 & -30.38   & 15.51 &  1  & \num{5.7e-07} & 0.97 & -41.7  & 18.17  & 1.8 \\
    & Gamma  & 0.1461 & 0.99 &-32.0  & 3.51  & 4.35  & \num{3.8e-04} & 0.98 & -44.4  & 2.06  & 9.56  \\
    & Beta & \num{3.837e-04} & 0.93 & -16.7 & 7.3 & 0.95  & \num{3.837e-04} & 0.98 & -24.74  & 6.37  & 0.64  \\
    & Log Logistic  & 0.70 & 0.99 &  -33.79   & 15.5  & 4.04  & 0.878 & 0.98 & -47.4 & 21.7  & 6.46  \\ \hline
    
    \end{tabular}
\end{table*}

When we examine both visualizations and goodness-of-fit statistics in this analysis, it is challenging to effectively fit distributions for locations that have low numbers of data points/paths, such as LOS-TUAS, NLOS-Sello, and NLOS-Airport. The p-values associated with most distributions in these locations exceed the significance threshold, indicating insufficient evidence to reject the null hypothesis of goodness of fit. However, visual inspection suggests that many distributions do not fit the data well in these locations. This discrepancy highlights the limitations of relying solely on statistical measures, especially in cases where the dataset size is small. 

Nevertheless, considering the entirety of visualizations and goodness-of-fit statistics reveals that several distributions, including log-normal, Nakagami, gamma, and beta distributions, exhibit satisfactory conformity with the LOS scheme. Notably, these distributions demonstrate acceptable values for both p and R values, indicating their suitability for modeling LOS propagation characteristics when a considerable number of data points are available. Conversely, in the case of the NLOS scheme, the data aligns well with the loglogistic distribution, particularly when a significant number of data points are available. This suggests that the loglogistic distribution may offer the most accurate representation of NLOS propagation characteristics under conditions where ample data points are present.

\subsection{Normalized Delay Distribution }

\begin{figure} [t]
    \centering
  \subfloat[Sello]{%
       \includegraphics[width=0.5\linewidth]{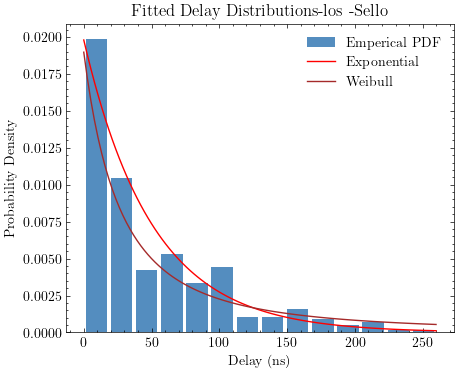}}
    \hfill
  \subfloat[Airport]{%
        \includegraphics[width=0.5\linewidth]{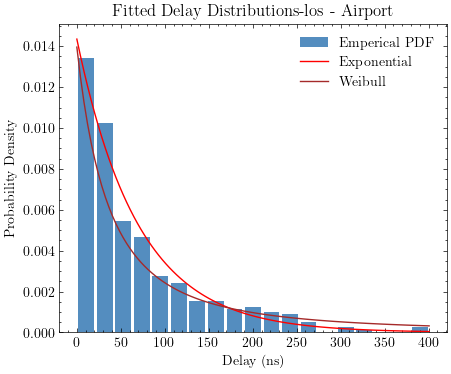}}
    \\
  \subfloat[TUAS]{%
        \includegraphics[width=0.5\linewidth]{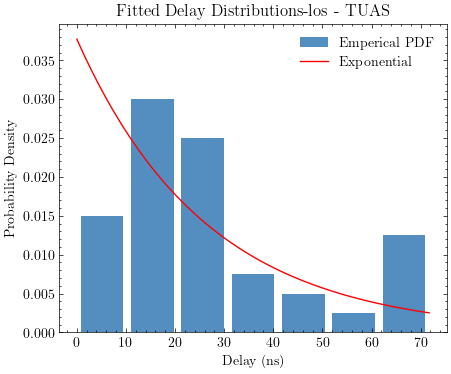}}
    \hfill
  \subfloat[TUAS2]{%
        \includegraphics[width=0.5\linewidth]{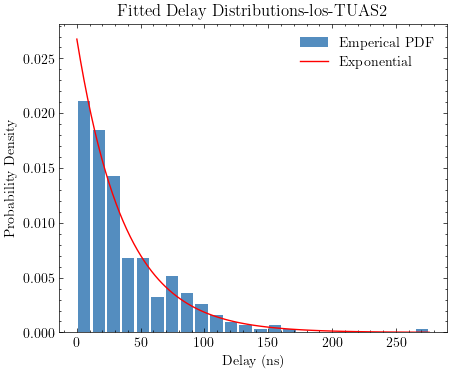}}
  \caption{Normalized Delay Distribution - Line-of-Sight}
  \label{fig:NDD - LOS} 
\end{figure}

\begin{figure} [t]
    \centering
  \subfloat[Sello]{%
       \includegraphics[width=0.5\linewidth]{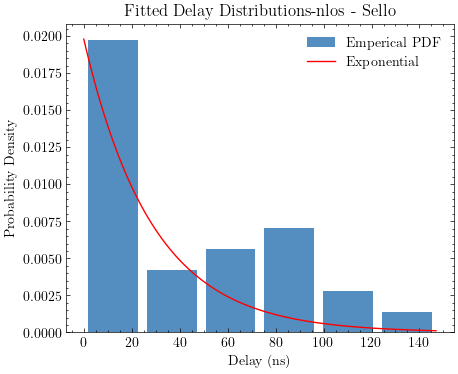}}
    \hfill
  \subfloat[Airport]{%
        \includegraphics[width=0.5\linewidth]{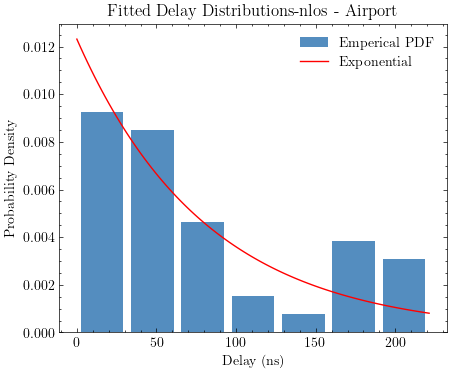}}
    \\
  \subfloat[TUAS]{%
        \includegraphics[width=0.5\linewidth]{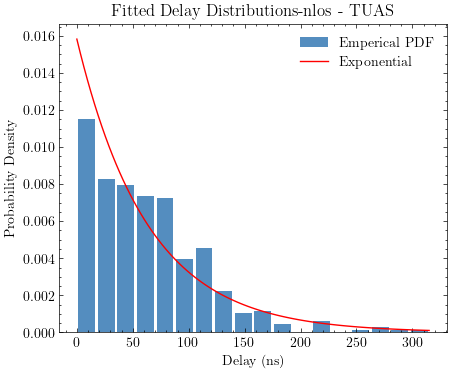}}
    \hfill
  \subfloat[TUAS2]{%
        \includegraphics[width=0.5\linewidth]{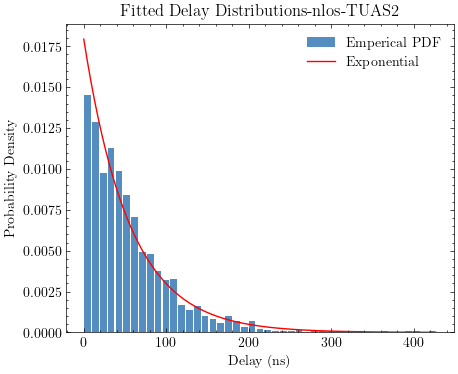}}
  \caption{Normalized Delay Distribution - Non-Line-of-Sight}
  \label{fig:NDD - NLOS} 
\end{figure}

To anticipate the delay distribution of the multipath components at D band we have plotted the PDFs related to the normalized delay as stated in (\ref{equ: delay_norm})  and compared the results with different theoretical PDFs. The resulting graphs are presented in Figures \ref{fig:NDD - LOS} and \ref{fig:NDD - NLOS}; the parameters are presented in Table \ref{tab:NDD - Results - Summary}. Since the lowest delay encounter from the normalized delay distribution is zero we have forced each distribution to locate at zero.


\begin{table} [t]
    \centering
    \caption{NDD - Results - Summary}
    \label{tab:NDD - Results - Summary}
    \resizebox{\columnwidth}{!}{\begin{tabular}{l |l| cccc| cccc}
    \hline
    \multirow{2}{*}{Location} &
    \multirow{2}{*}{Distribution} &
        \multicolumn{4}{c}{LOS} & 
        \multicolumn{4}{c}{NLOS} \\ 
    & &  p-value & R & Loc &  Scale  &  p-value & R &  Loc &  Scale \\ \hline
   \multirow{2}{*}{Sello} &  Exponential & 0.0401 & 0.99 &  0 &  50.52 & 0.562 & 0.99  & 0  & 43.51    \\
    & Weibull &  0.0142 & 0.99 & 0&  52.72 & \num{4.3e-22} & 0.91& 0  & 1.05  \\ \hline
    
    \multirow{2}{*}{Airport} & Exponential & 0.21 & 0.99 &  0 &  69.7  & 0.800 & 0.99 & 0 & 81.21 \\
    & Weibull & 0.09 & 0.99 & 0& 71.65  & \num{2.46e-66} & 0.90 & 0   & 1.05  \\ \hline
    
    \multirow{2}{*}{TUAS}  & Exponential & 0.161 & 0.99 &  0 &  26.5  & 0.002 & 0.99 & 0 & 63.2  \\
    & Weibull & \num{4.87e-32} & 0.91 & 0 & 1.05  & 0 & 0.99 & 0   & 1.05   \\ \hline
    
    \multirow{2}{*}{TUAS2} & Exponential & 0.283 & 0.99 &  0 &  37.5  & 0.0001 & 0.99 & 0 & 55.7   \\
    & Weibull & \num{1e-262}& 0.99 & 0 & 1.05  & 0 & 0.99 & 0 & 1.05  \\ \hline
    \end{tabular}}
\end{table}

Similar to the challenges encountered in fitting the distributions to the NPD, the normalized delay distribution (NDD) case also presents corresponding difficulties, particularly when the number of data points is low. However, unlike the NPD, the NDD reveals a limited number of distributions suitable for empirical fitting. Specifically, the Weibull and exponential distributions emerge as potential candidates for modeling the NDD behavior. Upon comprehensive analysis of statistics and visualizations, it becomes evident that the exponential distribution emerges as the most suitable fit for both LOS and NLOS schemes within the NDD framework. It offers the most accurate representation of delay characteristics in both LOS and NLOS scenarios despite the challenges posed by limited dataset sizes

\subsection{Number of Paths }

\begin{figure} [t]
    \centering
  \subfloat[Sello]{%
       \includegraphics[width=0.5\linewidth]{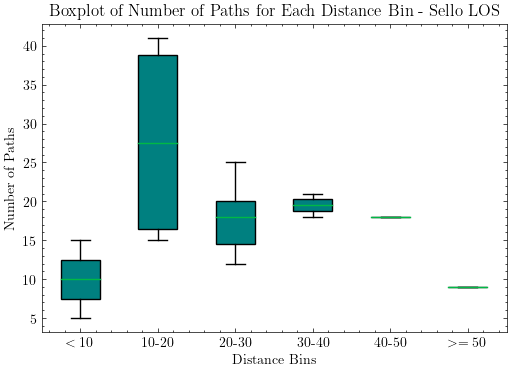}}
    \hfill
  \subfloat[Airport]{%
        \includegraphics[width=0.5\linewidth]{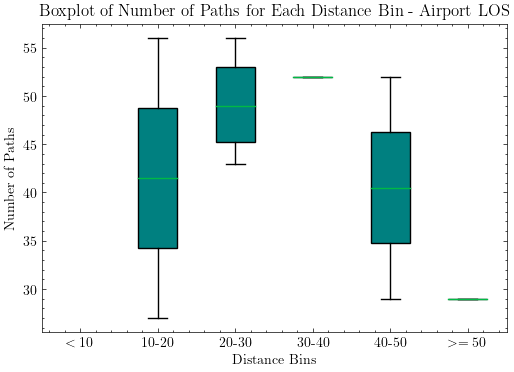}}
  \caption{Number of Paths - Line-of-Sight}
  \label{fig : NoP-LOS}
\end{figure}

\begin{figure} [t]
    \centering
  \subfloat[TUAS]{%
        \includegraphics[width=0.5\linewidth]{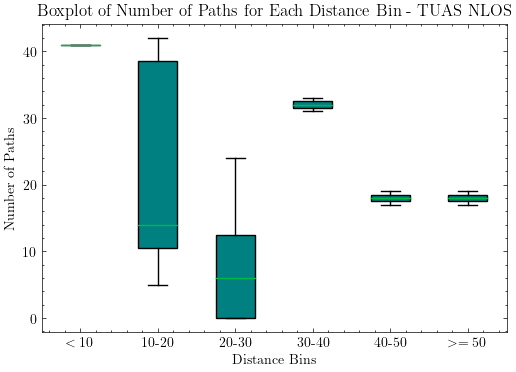}}
    \hfill
  \subfloat[TUAS2]{%
        \includegraphics[width=0.5\linewidth]{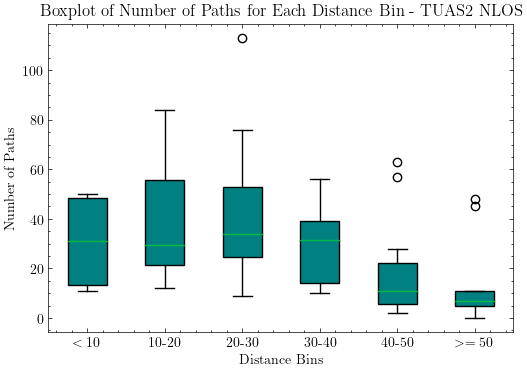}}
  \caption{Number of Paths - Non-Line-of-Sight}
  \label{fig : NoP-NLOS} 
\end{figure}

To discern the number of multipath components in D Band indoor wireless channels, we analyzed the number of paths (NoP) captured for each distance measurement between the Tx and the Rx. The visualizations related to this analysis are presented in Figures \ref{fig : NoP-LOS} and \ref{fig : NoP-NLOS}. For the LOS scenario, we considered the Sello and Airport locations, while TUAS and TUAS2 were selected for the NLOS scenario based on the highest number of data points available.


By examining the median values of each bin in the NLOS-TUAS2 location, which accounts for the highest number of data points, a clear pattern emerges. The median NoP increases up to the \qty{20}{m} -- \qty{30}{m} bin, following which it begins to decline with the Tx-Rx distance. However, considering the entire dataset, a consistent trend is evident. The median NoP peaks when the Tx-Rx distance falls within the \qty{10}{m} -- \qty{30}{m} range for both LOS and NLOS scenarios. This consistent observation underscores the importance of the \qty{10}{m}-\qty{30}{m} distance range in harboring the highest concentration of paths, which signifies critical propagation characteristics in both LOS and NLOS scenarios.

\subsection{Discussion}

Upon reviewing the results, it becomes apparent that multiple distributions can effectively characterize the NPD behavior of LOS multipath components, as observed in Sello and Airport locations, which have the highest numbers of data points related to LOS. However, in the case of NLOS multipaths, the log-logistic distribution demonstrates superior performance, particularly in locations TUAS and TUAS2 where the dataset size is substantial. Regarding the NDD, only two distributions, namely the Weibull and exponential, were viable ones for comparison. Notably, the exponential distribution emerges as the more suitable choice for both LOS and NLOS scenarios compared to the Weibull distribution. However, the conclusiveness of this observation may be limited due to the limited number of data points available for analysis.

Furthermore, the discrepancies between the results of the KS test and Q-Q plots are noted. These inconsistencies may stem from the relatively low number of data points used in the analysis. The KS test's sensitivity to dataset size can lead to unreliable performance when data points are scarce, while Q-Q plots may tend to overfit small datasets. In conclusion, a larger dataset size would facilitate more accurate conclusions and mitigate the influence of data point limitations on statistical tests and analyses

\section{Conclusions }\label{sec:conclusion} 

In this paper, we analyzed the spatial-temporal characteristics of channel propagation in the D band. Specifically, detailed analysis was conducted on power, delay, and the number of paths, considering both LOS and NLOS scenarios. Theoretical distributions were utilized to characterize the behavior of multipath components for power and delay in the D band. While several distributions accurately characterize the behavior of the LOS power distribution, other distributions, namely the NLOS power distribution characterized by the log-logistic distribution, as well as the LOS and NLOS delay distributions characterized by the exponential distribution, exhibit specific characteristics that can only be accurately captured by these relevant distributions. To further augment our research endeavors, we intend to derive location-specific statistical distributions of the angle of arrival/departure as well. These distributions, coupled with a predetermined antenna array configuration, will facilitate the generation of random MIMO channels by arbitrarily summing paths from the derived distributions. This approach will enable the generation of random yet realistic channels for indoor MIMO communications in the D band, thereby advancing our understanding and capability in this domain.

\section*{Acknowledgment}

This work was supported by the European Union Smart Networks and Services Joint Undertaking (SNS JU) under grant agreement no. 101096949 (TERA6G). This research was also supported by the Research Council of Finland (former Academy of Finland) 6G Flagship Programme (Grant Number: 346208).

\bibliographystyle{IEEEtran}
\bibliography{conf_short_1,jour_short_1,jour_full,conf_full,references}

\end{document}